\def\be{\begin{equation}}
\def\ee{\end{equation}}
\def\ba{\begin{array}}

\def\ea{\end{array}}

\documentclass[12pt]{article}
\usepackage{amsfonts,,amsmath}
\begin{document}
\parskip=3pt
\parindent=18pt
\baselineskip=20pt
\setcounter{page}{1}
\centerline{\large\bf General SIC-Measurement Based Entanglement Detection}
\vspace{6ex}
\centerline{{\sf Bin Chen,$^\star$}
\footnote{\sf chenbin5134@163.com}
~~~{\sf Tao Li,$^\star$}
\footnote{\sf lt881122@sina.com}
~~~and~~~ {\sf Shao-Ming Fei $^{\star,\dag}$}
\footnote{\sf feishm@cnu.edu.cn}
}
\vspace{4ex}
\centerline
{\it $^\star$ School of Mathematical Sciences, Capital Normal University, Beijing 100048, China}\par
\centerline
{\it $^\dag$ Max-Planck-Institute for Mathematics in the Sciences, 04103 Leipzig, Germany}\par
\vspace{6.5ex}
\parindent=18pt
\parskip=5pt
\begin{center}
\begin{minipage}{5in}
\vspace{3ex}
\centerline{\large Abstract}
\vspace{4ex}
We study the quantum separability problem by using general symmetric informationally complete measurements
and present separability criteria for both $d$-dimensional bipartite and multipartite systems. The criterion for
bipartite quantum states are effective in detecting several well known classes of quantum states. For isotropic states, it becomes both
necessary and sufficient. Furthermore, our criteria can be experimentally implemented and the criterion for two-qudit states
requires less local measurements than the one based on mutually unbiased measurements.
\end{minipage}
\end{center}

\newpage

\section{Introduction}
One of the most fundamental and intriguing tasks in quantum information theory and quantum information
processing is the detection of entanglement. It is widely known that entangled states are useful resources
in many quantum cryptography protocols, and can be used to enhance efficiency of quantum computing
(see reviews \cite{Horo,g09} and the references therein). There have been numerous criteria to distinguish
quantum entangled states from the separable ones, such as positive partial transposition criterion \cite{p96,ho96,ho97},
realignment criterion \cite{r1,r2,r3,r4,r5}, covariance matrix criterion \cite{c1}, and correlation matrix
criterion \cite{c2,c3}. In Ref. \cite{li14}, Li \emph{et al}. proposed a generalized form of the correlation
matrix criterion which is more effective than the previous criteria.

While mathematical methods presented above have been extensively studied, experimental implementation of
entanglement detection for unknown quantum states
has fewer results \cite{Bell,yusxia,liming,zmj}. In Ref. \cite{spe12}, the authors linked the separability
problem with the concept of mutually unbiased bases (MUBs) \cite{Woo}. They presented separability criteria
for two-qudit, multipartite and continuous-variable quantum systems. These separability criteria are shown to
be very powerful, and can be implemented experimentally. After that, Chen \emph{et al}. \cite{chenbin} generalized
such idea and provided a separability criterion for two-qudit states by using mutually unbiased measurements (MUMs) \cite{Kal}.
It is shown that the criterion based on MUMs is more effective than the criterion based on MUBs,
and for isotropic states this criterion becomes both necessary and sufficient. Very recently, Liu \emph{et al}. \cite{Liu} derived
separability criteria for arbitrary high-dimensional bipartite and multipartite systems using sets of MUMs.

Besides mutually unbiased bases, another related topic in quantum information theory is the symmetric
informationally complete positive operator-valued measures (SIC-POVMs). MUBs and SIC-POVMs have many interesting
and useful connections, from both operational link \cite{bened} and applications in quantum information theory such as
quantum state tomography \cite{Woo,qst1,qst2,Renes} and uncertainty relations \cite{RastEur}. In \cite{Appleby}, the author
introduced the concept of general symmetric informationally complete (SIC) measurements in which the elements need not be of rank
one, and showed that such general SIC measurements exist in all finite dimensions. Recently, Gour and Kalev  \cite{Gour} constructed
the set of all general SIC measurements from the generalized Gell-Mann matrices. Naturally, one may expect that these general symmetric
informationally complete measurements can be also used to detect entanglement experimentally.

In this paper we present separability criteria for both $d$-dimensional bipartite and multipartite systems
by using the general symmetric informationally complete measurements. The criterion for bipartite quantum states are shown
to be powerful in detecting some well known classes of quantum states. Moreover, this criterion requires less local measurements 
than the criterion based on mutually unbiased measurements.
The paper is organized as follows. In Section 2, we recall some basic notions of SIC-POVMs and the general
symmetric informationally complete measurements. In Section 3, we provide separability criteria based on the
general symmetric informationally complete measurements, and illustrate the power of entanglement detection
via some examples. We also compare the criterion for bipartite systems to the one based on mutually unbiased measurements.
We conclude the paper in Section 4.

\section{SIC-POVMs and General SIC Measurements}
Let us first review some basic definitions of SIC-POVMs and general symmetric informationally
complete measurements. A POVM with $d^{2}$ rank one operators acting on $\mathbb{C}^{d}$ is symmetric
informationally complete, if every operator is of the form
$$
P_{j}=\frac{1}{d}|\phi_{j}\rangle\langle\phi_{j}|,~~~j=1,2,\ldots,d^{2},
$$
the vectors $|\phi_{j}\rangle$ satisfying
$$
\mid\langle\phi_{j}|\phi_{k}\rangle\mid^{2}=\frac{1}{d+1},~~~j\neq k.
$$
The existence of SIC-POVMs in arbitrary dimension $d$ is an open problem. Only in a number of low-dimensional cases,
the existence of SIC-POVMs has been proved analytically, and numerically for all dimensions up to 67 (see \cite{Scott} and the references therein).

Recently, the concept and different constructions of general SIC measurements were introduced in Ref. \cite{Appleby,Gour}.
A set of $d^{2}$ positive-semidefinite operators $\{P_{\alpha}\}_{\alpha=1}^{d^{2}}$ on $\mathbb{C}^{d}$ is said to be a general SIC measurements, if \\
\indent(1) $\sum_{\alpha=1}^{d^{2}}P_{\alpha}=I$,\\
\indent(2) $\mathrm{Tr}(P_{\alpha}^{2})=a,~ \mathrm{Tr}(P_{\alpha}P_{\beta})=\frac{1-da}{d(d^{2}-1)},~\forall\alpha,\beta\in\{1,2,\ldots,d^{2}\},~\alpha\neq\beta$,\\
where $I$ is the identity operator, the parameter $a$ satisfies $\frac{1}{d^{3}}<a\leq\frac{1}{d^{2}}$,
$a={1}/{d^{2}}$ if and only if all $P_{\alpha}$ are rank one, which gives rise to a SIC-POVM.
It can be shown that $\mathrm{Tr}(P_{\alpha})=\frac{1}{d}$ for all $\alpha$ \cite{Gour}.

Like the mutually unbiased measurements, these general symmetric informationally complete measurements can be
also explicitly constructed for arbitrary dimensional spaces \cite{Gour}. Let $\{F_{\alpha}\}_{\alpha=1}^{d^{2}-1}$
be a set of $d^{2}-1$ Hermitian, traceless operators acting on $\mathbb{C}^{d}$,
satisfying $\mathrm{Tr}(F_{\alpha}F_{\beta})=\delta_{\alpha,\beta}$. Define $F=\sum_{\alpha=1}^{d^{2}-1}F_{\alpha}$, then the $d^{2}$ operators \\
\begin{equation}
\begin{split}
P_{\alpha}&=\frac{1}{d^{2}}I+t[F-d(d+1)F_{\alpha}],~~~\alpha=1,2,\ldots,d^{2}-1,\\
P_{d^{2}}&=\frac{1}{d^{2}}I+t(d+1)F,
\end{split}
\end{equation}
form a general SIC measurements. Here $t$ should be chosen such that $P_{\alpha}\geq0$, and the parameter $a$ is given by
\be\label{pp}
a=\frac{1}{d^{3}}+t^{2}(d-1)(d+1)^{3}.
\ee

These general symmetric informationally complete measurements have many useful applications in quantum information
theory. In Ref. \cite{RastSIC}, based on the calculation of the so-called index of coincidence, the author derived a
number of uncertainty relation inequalities via general SIC measurements. In the following, we study
entanglement detection using general symmetric informationally complete measurements.

\section{General SIC-POVM Based Separability Criterion}
The entanglement detection via SIC-POVMs has been briefly discussed in Ref. \cite{RastEur}.
But the method is subject to the existence of SIC-POVMs, which is an open question. Fortunately, unlike the
SIC-POVMs, general symmetric informationally complete measurements do exist for arbitrary dimension $d$,
and the separability criterion for two-qudit states can be explicitly presented.

\emph{Theorem 1}. Let $\rho$ be a density matrix in $\mathbb{C}^{d}\bigotimes\mathbb{C}^{d}$. Let
$\{P_{j}\}_{j=1}^{d^{2}}$ and $\{Q_{j}\}_{j=1}^{d^{2}}$ be any two sets of general symmetric informationally
complete measurements on $\mathbb{C}^{d}$ with the same parameter $a$.
Define $J_{a}(\rho)=\sum_{j=1}^{d^{2}}\mathrm{Tr}(P_{j}\bigotimes Q_{j}\rho)$. If $\rho$ is separable,
then $J_{a}(\rho)\leq\frac{ad^{2}+1}{d(d+1)}$.

\emph{Proof}. It is obvious that $J_{a}(\rho)$ is a linear function of $\rho$. Hence we need only to consider pure
separable states, $\rho=|\phi\rangle\langle\phi|\otimes|\psi\rangle\langle\psi|$. We have
\begin{eqnarray*}
J_{a}(\rho) & = & \sum_{j=1}^{d^{2}}\mathrm{Tr}(P_{j}\bigotimes Q_{j}\rho)\\
& = & \sum_{j=1}^{d^{2}}\mathrm{Tr}(P_{j}|\phi\rangle\langle\phi|)\mathrm{Tr}(Q_{j}|\psi\rangle\langle\psi|)\\
& \leq &
\frac{1}{2}\sum_{j=1}^{d^{2}}\{[\mathrm{Tr}(P_{j}|\phi\rangle\langle\phi|)]^{2}+[\mathrm{Tr}(Q_{j}|\psi\rangle\langle\psi|)]^{2}\}.
\end{eqnarray*}

Note that
$$
\sum_{j=1}^{d^{2}}[\mathrm{Tr}(P_{j}\rho)]^{2}=\frac{(ad^{3}-1)\mathrm{Tr}(\rho^{2})+d(1-ad)}{d(d^{2}-1)}
$$
for any density matrix $\rho$ in $\mathbb{C}^{d}$ \cite{RastSIC}, and $\mathrm{Tr}(\rho^{2})=1$ as $\rho$ is a pure state.
Thus we have $J_{a}(\rho)\leq\frac{ad^{2}+1}{d(d+1)}$. \quad $\Box$

Let $\{P_{j}\}_{j=1}^{d^{2}}$ be a set of general symmetric informationally complete measurements
on $\mathbb{C}^{d}$ with the parameter $a$. Let $\overline{P_{j}}$ denote the conjugation of $P_{j}$.
Then $\{\overline{P_{j}}\}_{j=1}^{d^{2}}$ is another set of general SIC-POVM with the same parameter $a$.
To show effectiveness of our criterion, let us consider some examples in the following.

\emph{Example 1.} We first consider the maximally entangled pure state $|\Phi^{+}\rangle=\frac{1}{\sqrt{d}}\sum_{i=0}^{d-1}|ii\rangle$. We have
\begin{eqnarray*}
J_{a}(|\Phi^{+}\rangle) & = & \sum_{j=1}^{d^{2}}\mathrm{Tr}(P_{j}\bigotimes \overline{P_{j}}|\Phi^{+}\rangle\langle\Phi^{+}|)\\
& = & da >\frac{ad^{2}+1}{d(d+1)},
\end{eqnarray*}
since $a>\frac{1}{d^{3}}$ from (\ref{pp}). Thus the criterion can detect all the maximally entangled pure states.

In Ref. \cite{RastEur}, the maximally entangled pure states can be also detected by the criterion based on the SIC-POVMs,
but the criterion depends on the existence of SIC-POVMs. Here we can detect all the maximally entangled pure states for arbitrary dimensions.

\emph{Example 2.} Let us consider the isotropic states, which are locally unitarily equivalent to a maximally entangled state mixed with white noise:
$$
\rho_{iso}=\alpha|\Phi^{+}\rangle\langle\Phi^{+}|+\frac{1-\alpha}{d^{2}}I,
$$
where $0\leq\alpha\leq1.$ One can easily get that
\begin{eqnarray*}
J_{a}(\rho_{iso}) & = & \sum_{j=1}^{d^{2}}\mathrm{Tr}(P_{j}\bigotimes \overline{P_{j}}\rho_{iso})\\
& = & da\alpha +\frac{1-\alpha}{d^{2}}.
\end{eqnarray*}
If $\alpha>\frac{1}{d+1}$, then $J_{a}(\rho_{iso})>\frac{ad^{2}+1}{d(d+1)}$ and $\rho_{iso}$ must be entangled by our theorem.
Thus the criterion can detect all the entanglement of the isotropic states, since it has been proven that $\rho_{iso}$ is
entangled for $\alpha>\frac{1}{d+1}$, and separable for $\alpha\leq\frac{1}{d+1}$ \cite{ber05}. That is to say that
our criterion is both necessary and sufficient for the separability of isotropic states,
similar to the criterion based on mutually unbiased measurements \cite{chenbin}.

\emph{Example 3.} We consider now the Bell-diagonal states,
\begin{equation*}
    \rho_{Bell}=\sum_{s,t=0}^{d-1}p_{s,t}|\Phi_{s,t}^{+}\rangle\langle\Phi_{s,t}^{+}|,
\end{equation*}
where $p_{s,t}\geq 0$, $\sum_{s,t=0}^{d-1}p_{s,t}=1$, $|\Phi_{s,t}^{+}\rangle=(U_{s,t}\bigotimes I)|\Phi^{+}\rangle$,
and $U_{s,t}=\sum_{j=0}^{d-1}\zeta_{d}^{sj}|j\rangle\langle j\oplus t|$, $s,t=0,1,\cdots,d-1,$ are Weyl operators,
$\zeta_{d}=e^{\frac{2\pi\sqrt{-1}}{d}}$ and $j\oplus t$ denotes $(j+t)$ mod $d$.
Denoting $A\geq B$ if $A-B$ is positive for operators $A$ and $B$, we have
$$
\rho_{Bell}\geq c|\Phi_{c}\rangle\langle\Phi_{c}|,
$$
where $c=\mathrm{max}\{p_{s,t}:s,t=0,1,\cdots,d-1\}$, $\frac{1}{d^{2}}\leq c\leq 1$, and $|\Phi_{c}\rangle$
is the corresponding maximally entangled pure state. Thus we obtain
\begin{eqnarray*}
J_{a}(\rho_{Bell}) & = & \sum_{j=1}^{d^{2}}\mathrm{Tr}(P_{j}\bigotimes \overline{P_{j}}\rho_{Bell})\\
& \geq & \sum_{j=1}^{d^{2}}\mathrm{Tr}[(P_{j}\bigotimes \overline{P_{j}}(c|\Phi_{c}\rangle\langle\Phi_{c}|)]\\
& = & cda.
\end{eqnarray*}
If $c>(1+\frac{1}{ad^{2}})/(d+1)$, then
$J_{a}(\rho_{Bell})>\frac{ad^{2}+1}{d(d+1)}$ and $\rho_{Bell}$ must be entangled by the theorem. It can be easily seen
that the criterion detects more entanglement as $a$ increases.

When $a=\frac{1}{d^{2}}$, i.e. the general SIC measurements
is given by the rank one SIC-POVM, we get $c>\frac{2}{d+1}$. This condition is the same as the one obtained by
the mutually unbiased measurements, in which the parameter $\kappa=1$ \cite{chenbin}, i.e. the mutually unbiased measurements
is given by the mutually unbiased bases. But for general cases, we do not know which criterion is more effective in detecting
the Bell-diagonal states, since we do not know the relation between the parameter $a$ and $\kappa$ for fixed $d$.

\emph{Example 4.} In \cite{yusx}, the authors discussed the entanglement of a bipartite state in $\mathbb{C}^{d}\bigotimes\mathbb{C}^{d}$:
\begin{equation*}
    \rho=a_{1}|\Phi^{+}\rangle\langle\Phi^{+}|+\sum_{k=1,i=2}^{d}\frac{a_{i}}{d}|k\rangle\langle k|\otimes|k+i-1\rangle\langle k+i-1|,
\end{equation*}
where $a_{i}>0, i=1,2,\ldots,d, \sum_{i=1}^{d}a_{i}=1.$ If $a_{i}\geq a_{1}(i\neq1),$ then the state is separable \cite{yusx}. We now consider a special case where $a_{i}=a_{2},$ for $i\neq1$. It is obvious that $\rho$ must be entangled, when $a_{1}>\frac{1}{d}$. We employ two sets of general SIC measurements $\{P_{j}\}_{j=1}^{d^{2}}$ and $\{\overline{P_{j}}\}_{j=1}^{d^{2}}$ as above, and denote the diagonal elements of $P_{j}$ as $\{P_{j}^{(1)},P_{j}^{(2)},\ldots,P_{j}^{(d)}\}$. Note that $\sum_{k=1}^{d}[P_{j}^{(k)}]^{2}\leq[\mathrm{Tr}(P_{j})]^{2}$, since $P_{j}^{(k)}$'s are non-negative numbers. Then we have
\begin{eqnarray*}
J_{a}(\rho) & = & \sum_{j=1}^{d^{2}}\mathrm{Tr}(P_{j}\bigotimes \overline{P_{j}}\rho)\\
& = & a_{1}da+\frac{a_{2}}{d}\sum_{j=1}^{d^{2}}\sum_{k=1,i=2}^{d}P_{j}^{(k)}P_{j}^{(k+i-1)}\\
& = & a_{1}da+\frac{a_{2}}{d}\sum_{j=1}^{d^{2}}\sum_{k=1}^{d}P_{j}^{(k)}[\mathrm{Tr}(P_{j})-P_{j}^{(k)}]\\
& = & a_{1}da+\frac{a_{2}}{d}\sum_{j=1}^{d^{2}}\{[\mathrm{Tr}(P_{j})]^{2}-\sum_{k=1}^{d}[P_{j}^{(k)}]^{2}\}\\
& \geq & a_{1}da.
\end{eqnarray*}
Similar to the discussion of Bell-diagonal states, we can conclude that the state
$\rho=a_{1}|\Phi^{+}\rangle\langle\Phi^{+}|+\frac{a_{2}}{d}\sum_{k=1,i=2}^{d}|k\rangle\langle k|\otimes|k+i-1\rangle\langle k+i-1|$ is entangled when $a_{1}>(1+\frac{1}{ad^{2}})/(d+1)$. Note that when $d$ is large enough, the separability threshold $(1+\frac{1}{ad^{2}})/(d+1)$ derived from our criterion approaches to $\frac{1}{d}$, independent of the exact value of $a$.

The criterion for two-qudit states can also be extended to $d$-dimensional multipartite states. We have the following theorem.

\emph{Theorem 2}. Let $\rho$ be a density matrix in $(\mathbb{C}^{d})^{\otimes N}$. Let
$\{P_{j}^{(i)}\}_{j=1}^{d^{2}}$, $i=1,2,\ldots,N,$ be $N$ sets of general symmetric informationally
complete measurements on $\mathbb{C}^{d}$ with the parameters $a_{i}$, respectively.
If $\rho$ is fully separable, then
$$
J(\rho)\leq\frac{1}{N}\sum_{i=1}^{N}\frac{a_{i}d^{2}+1}{d(d+1)},
$$
where $J(\rho)=\sum_{j=1}^{d^{2}}\mathrm{Tr}(\bigotimes_{i=1}^{N}P_{j}^{(i)}\rho)$.

\emph{Proof}. Similar to $J_{a}(\rho)$ defined in Theorem 1, $J(\rho)$ is also a linear function of $\rho$.
Therefore we only need to consider fully separable pure states, $\rho=\bigotimes_{i=1}^{N}|\phi_{i}\rangle\langle\phi_{i}|$. We have
\begin{eqnarray*}
J(\rho) & = & \sum_{j=1}^{d^{2}}\prod_{i=1}^{N}\mathrm{Tr}(P_{j}^{(i)}|\phi_{i}\rangle\langle\phi_{i}|).
\end{eqnarray*}
Note that $0\leq\mathrm{Tr}(P_{j}^{(i)}|\phi_{i}\rangle\langle\phi_{i}|)\leq\frac{1}{d}$. Using the inequality for $N$ non-negative real numbers \cite{Liu}:
$x_{1}x_{2}\cdots x_{N}\leq(\frac{x_{1}^{2}+x_{2}^{2}+\cdots+x_{N}^{2}}{N})^{\frac{N}{2}}$, we obtain
\begin{eqnarray*}
J(\rho) & \leq & \sum_{j=1}^{d^{2}}\{\frac{1}{N}\sum_{i=1}^{N}[\mathrm{Tr}(P_{j}^{(i)}|\phi_{i}\rangle\langle\phi_{i}|)]^{2}\}^{\frac{N}{2}}\\
& \leq & \sum_{j=1}^{d^{2}}\frac{1}{N}\sum_{i=1}^{N}[\mathrm{Tr}(P_{j}^{(i)}|\phi_{i}\rangle\langle\phi_{i}|)]^{2}\\
& = & \frac{1}{N}\sum_{i=1}^{N}\sum_{j=1}^{d^{2}}[\mathrm{Tr}(P_{j}^{(i)}|\phi_{i}\rangle\langle\phi_{i}|)]^{2}\\
& = & \frac{1}{N}\sum_{i=1}^{N}\frac{a_{i}d^{2}+1}{d(d+1)}.
\end{eqnarray*}
Therefore, $J(\rho)\leq\frac{1}{N}\sum_{i=1}^{N}\frac{a_{i}d^{2}+1}{d(d+1)}$ holds for all $d$-dimensional fully separable states $\rho$.  \quad $\Box$

Obviously, Theorem 1 is a special case of Theorem 2 for $N=2$ and $a_{1}=a_{2}$.

Let us make a comparison between the criterion for bipartite quantum states presented in this work and the one based on mutually unbiased measurements in
Ref. \cite{chenbin}. Let $\{P_{j}\}_{j=1}^{d^{2}}$ be a set of general
SIC measurements on $\mathbb{C}^{d}$ with the parameter $a$. By expanding a two-qudit state $\rho$ in terms of the operator basis
adopted in $\{P_{j}\}_{j=1}^{d^{2}}$, we get
\begin{eqnarray*}
J_{a}(\rho) & = & \sum_{j=1}^{d^{2}}\mathrm{Tr}(P_{j}\bigotimes P_{j}\rho)\\
& = & \frac{1}{d^{2}}+\frac{2(ad^{2}-1/d)}{d^{2}-1}\mathrm{Tr}(T),
\end{eqnarray*}
where $T$ is the correlation matrix of $\rho$. If $\rho$ is separable, then we obtain $\mathrm{Tr}(T)\leq\frac{d-1}{2d}$,
which is the same as the inequality deduced from the criterion based on the mutually unbiased measurements in Ref. \cite{chenbin}.
It is also a special case of the inequality satisfied by separable states \cite{li14}.
However, the separability criterion based on general SIC-Measurement and mutually unbiased measurement
can be experimentally implemented. One does not need to do tomography of an unknown state first.
Moreover, the criterion for bipartite quantum states based on general SIC measurements is superior to the one based on mutually unbiased measurements,
since the former only needs $d^{2}$ joint local measurements, while the later needs $d(d+1)$ joint local measurements,
which greatly reduces the experimental implementation complexity.

\section{Conclusion and Discussions}
We have studied the separability problem via general symmetric informationally complete measurements.
More experimentally feasible quantum separability criteria for two-qudit states and for $d$-dimensional multipartite states have been presented.
The criterion for bipartite quantum states has been shown to be powerful in detecting the quantum entanglement for maximally
entangled pure states, the isotropic states and the Bell-diagonal states. Especially, for isotropic states,
this criterion is both necessary and sufficient.
Comparing with the criterion based on the mutually unbiased measurements,
our criterion based on the general symmetric informationally complete measurements
requires much less measurements. It would be also worthwhile to generalize the present results to arbitrary high-dimensional multipartite systems.

\vspace{2.5ex}
\noindent{\bf Acknowledgments}\, \,
This work is supported by the NSFC under number 11275131.


\vspace{2.5ex}
\begin{thebibliography}{20}

\bibitem{Horo} R. Horodecki, P. Horodecki, M. Horodecki, and K. Horodecki, Rev. Mod. Phys. \textbf{81}, 865 (2009).

\bibitem{g09} O. Guhne and G. Toth, Phys. Rep. \textbf{474}, 1 (2009).

\bibitem{p96} A. Peres, Phys. Rev. Lett. \textbf{77}, 1413 (1996).

\bibitem{ho96} M. Horodecki, P. Horodecki, and R. Horodecki, Phys. Lett. A
\textbf{223}, 1 (1996).

\bibitem{ho97} P. Horodecki, Phys. Lett. A \textbf{232}, 333 (1997).

\bibitem{r1} O. Rudolph, Phys. Rev. A \textbf{67}, 032312 (2003).

\bibitem{r2} K. Chen and L. A. Wu, Quant. Inf. Comput. \textbf{3}, 193 (2003).

\bibitem{r3} M. Horodecki, P. Horodecki, and R. Horodecki, Open Syst. Inf.
Dyn. \textbf{13}, 103 (2006).

\bibitem{r4} K. Chen and L. A. Wu, Phys. Lett. A \textbf{306}, 14 (2002); Phys. Rev.
A \textbf{69}, 022312 (2004); P. Wocjan and M. Horodecki, Open Syst.
Inf. Dyn. \textbf{12}, 331 (2005).

\bibitem{r5} S. Albeverio, K. Chen, and S. M. Fei, Phys. Rev. A \textbf{68}, 062313
(2003).

\bibitem{c1} O. Guhne, P. Hyllus, O. Gittsovich, and J. Eisert, Phys. Rev.
Lett. \textbf{99}, 130504 (2007).

\bibitem{c2} J. D. Vicente, Quant. Inf. Comput. \textbf{7}, 624 (2007).

\bibitem{c3} J. D. Vicente, J. Phys. A: Math. Theor. \textbf{41}, 065309 (2008).

\bibitem{li14} M. Li, J. Wang, S. M. Fei, and X. Li-Jost, Phys. Rev. A \textbf{89}, 022325
(2014).

\bibitem{Bell} N. Gisin, Phys. Lett. A \textbf{154}, 201 (1991).

\bibitem{yusxia} S. Yu, J. W. Pan, Z. B. Chen and Y. D. Zhang, Phys. Rev. Lett. \textbf{91}, 217903 (2003).

\bibitem{liming} M. Li and S. M. Fei, Phys. Rev. Lett. \textbf{104}, 240502 (2010).

\bibitem{zmj} M. J. Zhao, T. Ma,  S. M. Fei and Z. X. Wang, Phys. Rev. A \textbf{83}, 052120 (2011).

\bibitem{spe12} C. Spengler, M. Huber, S. Brierley, T. Adaktylos, and B. C. Hiesmayr, Phys. Rev. A \textbf{86}, 022311 (2012).

\bibitem{Woo} W. K. Wootters and B. D. Fields, Ann. Phys. (N.Y.) \textbf{191}, 363 (1989).

\bibitem{chenbin} B. Chen, T. Ma, and S. M. Fei, Phys. Rev. A \textbf{89}, 064302 (2014).

\bibitem{Kal} A. Kalev and G. Gour, New J. Phys. \textbf{16}, 053038 (2014).

\bibitem{Liu} L. Liu, T. Gao, and F. l. Yan, arXiv: 1501.01717 [quant-ph] (2015).

\bibitem{bened} R. Beneduci, T. J. Bullock, P. Busch, C. Carmeli, T. Heinosaari, and A. Toigo, Phys. Rev. A \textbf{88}, 032312 (2013).

\bibitem{qst1} R. B. A. Adamson and A. M. Steinberg, Phys. Rev. Lett. \textbf{105}, 030406 (2010).

\bibitem{qst2} A. Fern$\acute{a}$ndez-P$\acute{e}$rez, A. B. Klimov, and C. Saavedra, Phys. Rev. A \textbf{83}, 052332 (2011).

\bibitem{Renes} J. M. Renes, R. Blume-Kohout, A. J. Scott, and C. M. Caves, J. Math. Phys. \textbf{45}, 2171 (2004).

\bibitem{RastEur} A. E. Rastegin, Eur. Phys. J. D \textbf{67}, 269 (2013).

\bibitem{Appleby} D. M. Appleby, Opt. Spectrosc. \textbf{103}, 416 (2007).

\bibitem{Gour} G. Gour and A. Kalev, J. Phys. A: Math. Theor \textbf{47} 335302 (2014).

\bibitem{Scott} A. J. Scott and M. Grassl, J. Math. Phys. \textbf{51}, 042203 (2010).

\bibitem{RastSIC} A. E. Rastegin, Phys. Scr. \textbf{89}, 085101 (2014).

\bibitem{ber05} R. A. Bertlmann, K. Durstberger, B. C. Hiesmayr, and P. Krammer, Phys. Rev. A \textbf{72}, 052331 (2005).

\bibitem{yusx} S. X. Yu and N. L. Liu, Phys. Rev. Lett \textbf{95}, 150504 (2005).

\end{thebibliography}
\end{document}